# Gaze-based Screening of Autistic Traits for Adolescents and Young Adults using Prosaic Videos

Karan Ahuja
Carnegie Mellon University
Pittsburgh, USA
kahuja@cs.cmu.edu

Abhishek Bose
Indian Institute of Information
Technology Guwahati
Guwahati, India
abose550@gmail.com

Mohit Jain
Microsoft Research
Banglore, India
mohja@microsoft.com

Kuntal Dey
IBM Research
New Delhi, India
kuntadey@in.ibm.com

Anil Joshi
IBM Research
New Delhi, India
anijoshi@in.ibm.com

Krishnaveni Achary
Tamana NGO
New Delhi, India
krishnaveni@tamana.ngo

Blessin Varkey
Tamana NGO
New Delhi, India
blessinvarkey@tamana.ngo

Chris Harrison
Carnegie Mellon University
Pittsburgh, USA
chris.harrison@cs.cmu.edu

Mayank Goel
Carnegie Mellon University
Pittsburgh, USA
mayankgoel@cmu.edu

## ABSTRACT

Autism often remains undiagnosed in adolescents and adults. Prior research has indicated that an autistic individual often shows atypical fixation and gaze patterns. In this short paper, we demonstrate that by monitoring a user's gaze as they watch commonplace (i.e., not specialized, structured or coded) video, we can identify individuals with autism spectrum disorder. We recruited 35 autistic and 25 non-autistic individuals, and captured their gaze using an off-the-shelf eye tracker connected to a laptop. Within 15 seconds, our approach was 92.5% accurate at identifying individuals with an autism diagnosis. We envision such automatic detection being applied during e.g., the consumption of web media, which could allow for passive screening and adaptation of user interfaces.

## CCS CONCEPTS

• **Human-centered computing** → **Ubiquitous and mobile computing**.

## KEYWORDS

Autism; gaze tracking; health sensing.

## 1 INTRODUCTION

Autism Spectrum Disorder (ASD) is a universal and often life-long neuro-developmental disorder [5]. Individuals with ASD often present comorbidities such as epilepsy, depression, and anxiety. In the United States, in 2014, 1 out of 68 people was affected by autism [6], but worldwide, the number of affected people drops to 1 in 160 [1]. This disparity is primarily due to underdiagnosis and unreported cases in resource-constrained environments. Wiggins *et al.* found that, in the US, children of color are under-identified with ASD [40]. Missing a diagnosis is not without consequences; approximately 26% of adults with ASD are under-employed, and are under-enrolled in higher education [1]. Hategan *et al.* [17] echo these findings and highlight aging with autism as an emerging public health phenomenon.

Unfortunately, ASD diagnosis is not straightforward and involves a subjective assessment of the patient's behavior. Because such assessments can be noisy and even non-existent in low-resource environments, many cases go unidentified. Many such cases remain undiagnosed even when the patient reaches adolescence or adulthood [8]; and, as noted by Fletcher-Watson *et al.* [14], there has been little work done on ASD detection for adolescents and young adults.

There is a need for an objective, low-cost, and ubiquitous approach to diagnose ASD. Autism is often characterized by symptoms such as limited interpersonal and social communication skills, and difficulty in face recognition and emotion interpretation [38]. When watching video media, these symptoms can manifest as reduced eye fixation [21], resulting in characteristic gaze behaviors [4]. Thus, we developed an approach to screen patients with ASD using their gaze behavior while they watch videos on a laptop screen. We used a dedicated eye tracker to record the participant's gaze. With data from 60 participants (35 with ASD and 25 without ASD), our algorithm demonstrates 92.5% classification accuracy after the participants watched 15 seconds of the video. We also developed a proof-of-concept regression model that estimates the severity of the condition

---

[1] https://www.who.int/news-room/fact-sheets/detail/autism-spectrum-disorders

and achieves a mean absolute error of 2.03 on the Childhood Autism Rating Scale (CARS) [35].

One of the most common approaches to identify individuals with ASD involves studying family home videos and investigating an infant's gaze and interactions with their families [34]. However, having an expert carefully inspect hours of home video is expensive and unscalable. Our approach is more accessible and ubiquitous as we can directly sense the gaze of the user they watch videos. Such sensing can be directly deployed on billions of smartphones around the world that are equipped with a front-facing camera. In our current exploration, we use a dedicated eye-tracker but achieving similar performance using an unmodified smartphone camera is not far-fetched. Our results demonstrate that passively tracking a user's gaze pattern while they watch videos on a screen can enable robust identification of individuals with ASD. Past work has used specially-created visual content to detect ASD [7, 18, 32], but getting large sets of the population to watch specific videos is hard. Thus, we focus on generic content and selected four prosaic video scenes as a proof of concept.

Our research team includes experienced psychologists to inform the study design and contextualize the performance of the final system. Although our gaze tracking approach cannot yet replace a clinical assessment, we believe it could be valuable for screening individuals passively, as they consume media content on computing devices (*e.g.*, YouTube, Netflix, in-game cut scenes). We believe our efforts in estimating condition severity is also an essential first step towards building an entirely automated, in-home screening, and condition management tool. With rapid advancements in gaze tracking on consumer devices (*e.g.*, Apple iPhone, HTC Vive), autism detection could be included on modern computing devices as a downloadable app or background feature, and potentially reduce the number of undiagnosed cases. Such a system could also track the efficacy of treatment and interventions. Additionally, ASD detection could be used to automatically adapt user interfaces, which has been shown to improve accessibility [13, 16].

**Contributions:** The main contribution of our work are as follows:

- To the best of our knowledge, it is the first study to investigate the use of commonplace, unstructured prosaic videos to screen for the risk of autism and its severity.
- We explore the effects of the duration and annotations (for the salient object of visual interest) for the video on the classification performance.

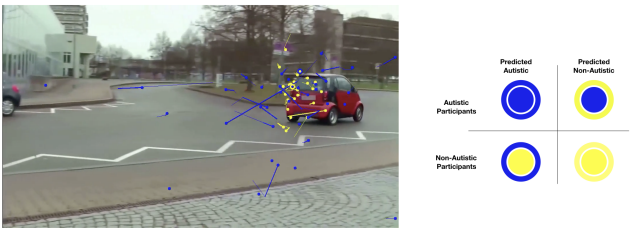

**Figure 1: Snapshot of participants' gaze on an example video still. Blue lines denote the gaze path of autistic participants, while yellow lines denote non-autistic participants. Classification legend on right.**

- The experiments are fully replicable as the gaze data and corresponding data analysis scripts will be made freely available.

## 2 BACKGROUND

Detecting ASD and analyzing the behavior of autistic individuals has been a long-standing area of research. The community has looked at autism diagnosis through various means, chiefly computational behavioral modeling [33], robot assistance [27] and sensor-driven methods [15]. Perhaps more importantly, researchers have shown that autistic users benefit from technology that adapts to the user's specific needs [24]. Our approach could help computers quickly screen users for the risk of autism and potentially adapt underlying interactions.

Autism is characterized by low attention towards social stimuli [11]. Experiments have established that autistic individuals tend to pay relatively lower attention to human behaviors such as hand gestures, faces, and voices, but concentrate more on non-social elements such as devices, vehicles, gadgets and other objects of "special interest" [12, 37]. Most of these studies have investigated the participant's gaze when exposed to controlled scenes, (*e.g.,* faces, objects against a plain background). Of late, researchers have started exploring what happens when participants are exposed to more natural stimuli, rather than an isolated face or object [9, 19, 31, 36, 41]. Seminal work by Klin *et al.* [25] utilized five video clips with rich social-content and showcased that adults with ASD are less likely to divert their attention to social stimuli, especially people's eyes. In follow-up work by Kemner *et al.* [22], it was found that such abnormalities are prominent in video, but not in still images. In recent years, Yaneva *et al.* [42] used gaze data from web searching tasks with annotated areas of interest to screen for autistic traits.

To make such diagnostic approaches scalable and generalizable, machine learning techniques have recently been applied. These approaches utilize eye tracking data on images [7, 20] or tailored videos [28]. Generalizability to unconstrained, generic media content, especially for adolescents and young adults, has not been explored previously.

## 3 DATA COLLECTION

We now explain our protocol for data collection. Our experimental setup consisted of a Tobii EyeX [3] connected to a Windows 10 laptop using USB 3.0 (Figure 2). Tobii EyeX records gaze data at a sampling rate of 60 Hz.

### 3.1 Participants

As mentioned previously, our target population are adolescents and young adults. For our participants with autism, we recruited from a special institution for individuals with ASD: 35 people (28 male, 7 female, age = 15-29 years). These participants were diagnosed with autism by clinical psychologists employing DSM-IV [26], followed by a functional assessment and CARS. We chose the DSM-IV over DSM-V [29] as many of the participants had already been evaluated under the DSM-IV protocol in the past. Therefore, rather than re-running them, we opted to use their existing scores to preserve homogeneity.

The participants with autism diagnoses had no other conditions of note. The autism severity ranged from mild to severe and was estimated using the CARS score [23], with the scores ranging between 30 to 39 (distribution shown in Table 2). See e.g., Mesibov *et al.* [30] for a discussion on how CARS can be used for adolescents and adults as well. For our non-autistic population as control, we recruited 25 participants (20 male, 5 female, age = 19-30) from the general public. There was an initial screening in place to ensure that none of the controls had autistic traits.

| Video Name | Duration (m:ss) | Description |
|---|---|---|
| Car Pursuit | 0:24 | Panning camera follows a red car while it was going through a roundabout |
| Dialog | 0:18 | Two persons talk to each other in front of the camera |
| Case Exchange | 0:26 | Various persons crossing the field of view while a text ribbon is showcased in the lower half of the screen |
| Ball Game | 0:26 | Three players with orange shirts and one player with a white shirt passes a ball around |

**Table 1: Brief descriptions of the video clips we used in our experiments. See also Video Figure.**

| CARS | 30 | 31 | 32 | 33 | 34 | 35 | 36 | 37 | 38 | 39 |
|---|---|---|---|---|---|---|---|---|---|---|
| N_sub | 3 | 5 | 6 | 4 | 5 | 7 | 3 | 0 | 1 | 1 |

**Table 2: Distribution of CARS score (ranging from 30 to 39) across our 35 autistic participants. We did not record CARS scores for our 25 non-autistic participants.**

## 3.2 Procedure

The study was conducted by a trained experimenter. For participants with an autism diagnosis, a clinical psychologist and school teacher were also present; mostly as observers and only provided assistance in rare cases of need. For our non-autistic participants, the clinical psychologist and the school teacher were not needed. Throughout the study, participants were asked to sit in a chair comfortably and look at a laptop's screen placed on a table in front of them.

The study started with a gaze calibration sequence provided by the Tobii EyeX's software. Participants were allowed to try the calibration step as many times as needed to get a tight registration. Three participants were not able to complete this calibration step, and were dropped from the study. Following the calibration, we asked the participants to watch four short videos in a random order.

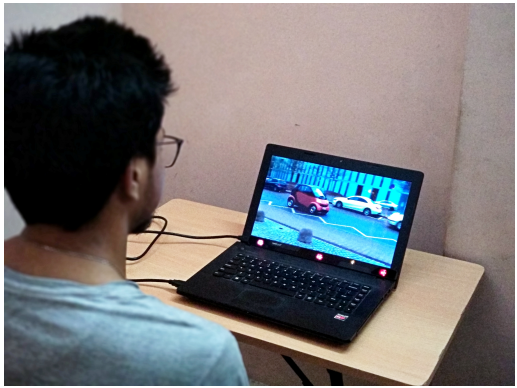

**Figure 2: Data collection setup. Gaze tracker is outlined in yellow.**

If their gaze shifted away from the screen for more than 500ms, the video playback automatically paused. When the participant's gaze returned to the screen, the video resumed. We logged all of the gaze movements ($x,y$ screen coordinates), along with timestamps, using a custom program running on the laptop. At the end of the study, we also recorded participant's demographic and CARS scores.

## 3.3 Exemplary Videos

We selected four prosaic videos from [2], a dataset containing dynamic stimuli with objects of interest annotated. A brief summary of these videos can be found in Table 1. Typically, the annotated object of interest corresponded to the visually salient stimuli. An example still frame from the dataset is showcased in Figure 3. The videos varied in length (from 18 to 26 seconds) and in difficulty with respect to the motion of salient objects.

## 4 METHODOLOGY

To develop generalizable features, we relied on following key observations by Wang *et al.* [39]:

- People with ASD have fewer fixations on the semantic objects of interest in the scene than on the background.
- People with ASD fixated at the semantic objects of interest significantly later than the control group, but not other objects.
- People with ASD had longer individual fixations than controls, especially for fixations on background.

Past work has leveraged a disjoint set of images to identify people with ASD. However, as Wang *et al.* and Klin *et al.* noted, there is a temporal component to an individual's fixation (such as delay and prolonged fixation). By using videos, instead of disjoint photos, we can model the temporal distribution of gaze points and track gaze changes in the same context and stimuli over time, rather than combining different responses to different stimuli. We calculate the following five features for each video sequence:

(1) **Standard deviation of gaze points:** We treat the gaze points across all the frames of the video as a distribution and then calculate its standard deviation.

(2) **Standard deviation of difference in gaze points:** We calculate the difference in gaze points in euclidean space between two consecutive frames, for all pairs in the video. We then calculate the standard deviation of this difference.

(3) **Standard deviation between the gaze and annotated object of interest:** We calculate the Manhattan distance between the point of gaze of the participant and the center of the box corresponding to the annotated object of interest for each frame. We then find the standard deviation of this distribution.

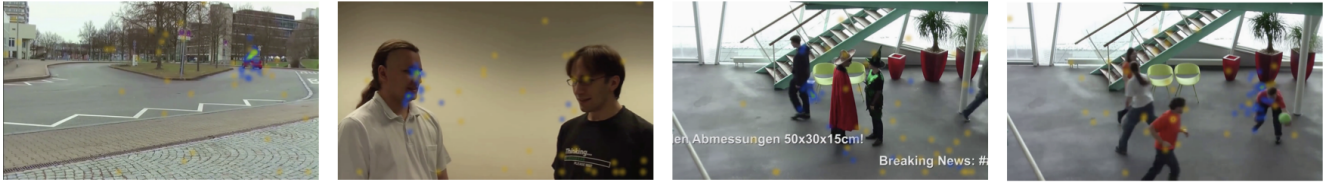

**Figure 3: Point cloud of eye-gaze locations on a sample frame from each video. Gaze of participants with an autism diagnosis ranges from blue to green (when points overlap), and yellow to red for non-autistic participants.**

(4) **RMSE between the gaze and annotated object of interest:** We treat the gaze points and the center of the annotated object of interest as two distributions and find the root mean square error between them.

(5) **Delay in looking at the object of interest:** We model the delay as the time difference between the first time the object of interest enters the scene and when the participant looks at it (i.e. when the point of gaze is inside the box corresponding to the annotated object of interest). We average all the delay values across multiple occurrences and multiple objects for each video.

While simple, these features are heavily motivated by prior research findings. Features 3 and 4 are motivated by the hypothesis that people with autism will focus on the background and not on the salient and semantic areas of interest. Feature 5 captures the delay with which people with autism may look at a salient object. Features 1 and 2 aim to capture differences in gaze patterns, even if there is no objects of interest present. Note that all these features, due to their temporal nature, can easily adapt to any sensor sampling rate (*i.e.*, the Tobii EyeX runs at 60 Hz, but the the same approach could also run on lower-sample-rate devices, such as a smartphone's front camera).

We employ these features to create a feature vector for each participant following each video. For autism *vs.* non-autism classification, we employ a Support Vector Machine with a third-degree polynomial kernel. For estimating severity of autism, we use CARS score as the output. We employ a multi-layer peceptron regressor (sklearn - default hyper-parameters), which performs well in extrapolating beyond values seen in our training data (e.g. useful when train data has values within 31-37, and test data is within 30-39).

## 5 RESULTS & DISCUSSION

We perform all of our analyses for our autism detection using cross validation. More specifically, given our test set of 60 participants, we run a 3 fold cross-validation and repeat this one hundred times (random sets, but keeping our test set balanced with autistic and non-autistic participants). This results in a 40 train and 20 test split of participants for each iteration, wherein no participant appears in both the splits. This helps us to evaluate the generalizability of our model and reduces overfitting. For autism severity estimation, we use leave one out cross validation as our evaluation metric due to the limited training data per CARS score (see Table 2).

Prior work has often used annotated objects of interest or custom content to make predictions [42]. Therefore, to study its trade-offs we test our approach with and without the annotated object of interest. Note, that the information regarding the annotated object of interest is to capture the salient object in the scene and is a meta-data that is invisible to the participants. It is only used by our machine learning module. Therefore, to understand the utility of object-of-interest annotations, we first compare the classification performance with and without annotations. We also study how performance changes as we add more videos to make the final inference. Next, we investigate the relationship between the duration of videos and performance to better understand how much video is needed to achieve usable performance. Finally, we provide preliminary results for estimating the severity of autism.

### 5.1 Autism detection with annotated object of interest

For this approach we use all 5 of our features as described in Section 4 and concatenate them into a feature vector. We thus evaluate the extent to which the extra features built upon the annotated object of interest (features 3, 4 and 5) help the model to learn the correlations between gaze and object saliency, thereby producing better predictions. For car pursuit, dialog, case exchange and ball game videos, we achieved an average classification accuracy of 91.41%, 93.44%, 94.34% and 91.93%, respectively. When we concatenate all four videos and combine their features into one feature vector, we achieve an accuracy of 98.3%. This result suggests that with roughly 100 seconds of data, our approach can be surprisingly accurate.

### 5.2 Autism detection without annotated object of interest

While we chose to use videos that were pre-annotated with objects of interest, this metadata is rarely available for in-the-wild videos. Thus, for our approach to be feasible in the real world, our algorithm would need to work on unannotated videos. For this, we investigated our algorithm's performance when using features bereft of any annotation information (*i.e.*, features 1 and 2 described earlier). When re-running our analysis, we found our approach was 91.16%, 91.74%, 86.79% and 90.91% accurate for car pursuit, dialog, case exchange, and ball game videos, respectively. When we concatenated the four videos, our algorithm achieved an accuracy of 93.3%. The accuracy drop of 5% can be attributed to the lack of object saliency data given to the machine learning model as compared to the earlier result. Nonetheless, the high accuracy demonstrates feasibility of our approach.

### 5.3 Duration Simulation

To assess the latency of our approach, it is important to investigate the effect of video length on classification performance. The most straightforward procedure would be to iteratively trim videos to see how classification accuracy changes. However, this does not take into account that a specific section of a video is significantly more informative than other parts. Thus, we generate random segments of videos of varying durations, and train-test our system only on gaze data from these periods.

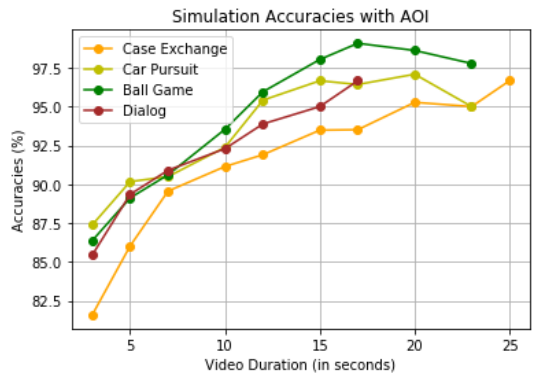

**Figure 4: Effect of video duration on classification accuracy when including features that leverage annotated objects-of-interest (AOI).**

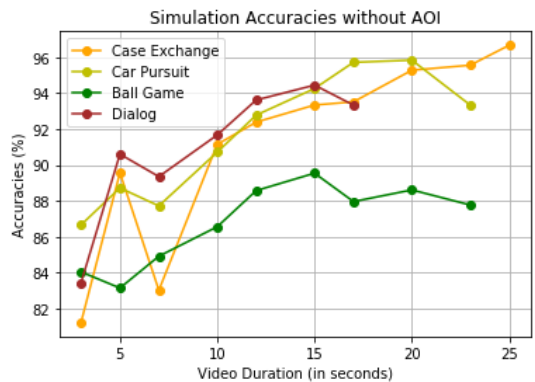

**Figure 5: Effect of video duration on classification accuracy when not using annotated objects-of-interest (AOI) data (*i.e.,*, raw video)**

When including features derived from object annotations, we notice that, as expected, the accuracy increases as training data increases (Figure 4). However, we note that the average accuracy is approximately 95.75% with just 15 seconds of video data when object of interest annotations are available.

Figure 5 plots accuracy assuming object of interest data is unavailable (offering a more generalizable result, applicable to any video). As before, we achieve strong accuracies — mean of 92.5% — in as little as 15 seconds of video.

### 5.4 Estimating Autism Severity

The capability to estimate severity of autism would be extremely valuable. Such a system can be used to perform longitudinal tracking of a person's condition, and understand the efficacy of treatment and different interventions. CARS score range from 15 to 60, with scores below 30 marking the non-autistic range, scores between 30 to 36.5 denoting mild to moderate autism, and above 37 denoting severe autism [10].

Our 35 autistic participants had CARS scores ranging from 30 to 39, which does not cover the full range (and our non-autistic participants had no known CARS scores). We used the available CARS scores to train our multi-layer peceptron regressor, which achieved a mean absolute CARS score error of 2.03, with a standard deviation of 1.37. Given the small participant count and CARS range, we note this result is promising but preliminary.

## 6 CONCLUSION

Autism affects people across many demographic groups and accurate diagnoses could help improve the quality of life for many people. Moreover, detecting severity of autism over time is an equally important task for many patients as it is a chronic condition. Our analysis shows that tracking a patient's gaze while they watch videos can be used to make screen patients with autistic traits. However, we show that such videos can be unstructured and need not be specifically prepared for target subjects. Moreover, such a system could potentially become pervasive as modern mobile devices ship with eye tracking capabilities. Our software could run in the background while users watch videos on the Internet or in some app. We believe that our effort is an important step in the direction of enabling completely passive, automatic identification and tracking of individuals with Autism Spectrum Disorder. Our results also suggest it may be possible to estimate the condition severity.